\documentclass{article}

 \usepackage{theorem}
 \theorembodyfont{\upshape}

\usepackage{amsmath,amssymb}
\newtheorem{theorem}{Theorem}

\newtheorem{lemma}{Lemma}
\newtheorem{corollary}{Corollary}
\newtheorem{remark}{Remark}

\numberwithin{equation}{section}

\newenvironment{acknowledgement}{\noindent{\textbf{ACKNOWLEDGEMENTS}}\par}

\DeclareMathOperator{\tr}{tr}
\DeclareMathOperator{\Ran}{Ran}

\renewcommand\H{\mathcal{H}}
\renewcommand\L{\mathrm{L}}
\newcommand\R{\mathbb R}
\newcommand\N{\mathbb N}

\newcommand\Z{\mathbb Z}

\newcommand\I{\mathcal{I}}

\newcommand\di{\mathrm d}

\newcommand\e{\mathrm{e}}

\newcommand\eps{\varepsilon}

\newcommand{\la}{\langle}
\newcommand{\ra}{\rangle}

\renewcommand\P{\mathbb P}
\newcommand\E{\mathbb E}
\newcommand\Pb{\mathbf P}
\newcommand\W{\mathbf W}

\newcommand{\norm}[1]{\left\lVert #1 \right\rVert}
\newcommand{\scal}[1]{\la #1 \ra}

\makeatletter
\newif\ifqed
\DeclareRobustCommand{\qed}{%
  \ifmmode 
  \else\leavevmode\unskip\penalty9999 \hbox{}\nobreak%
  \hspace{1.2em}\fi\hbox{\qedsymbol}\global\qedfalse}

\newcommand{\qedsymbol}{\rule[-0.2ex]{.9ex}{2ex}}

\newenvironment{proof}[1][]{\qedtrue\par
  \normalfont
  \topsep6\p@\@plus6\p@ \trivlist
  \item[\hskip1.45em\itshape
  \ifx @#1@ \textsf{{\itshape Proof}}\else\textsf{{\itshape #1}}\fi.\hskip.8em]\ignorespaces
}{%
  \ifqed\qed\fi\endtrivlist%
}
\makeatother

\begin{document}

\title
{New characterizations of the region of complete localization for random
 Schr\"odinger operators}
 

\author{Francois Germinet
 \thanks{Universit\'e de Cergy-Pontoise,
D\'epartement de Math\'ematiques,
Site de Saint-Martin,
2 avenue Adolphe Chauvin,
95302 Cergy-Pontoise cedex, France.
 Email: germinet@math.u-cergy.fr}
\and  {Abel Klein}
\thanks{University of California, Irvine,
Department of Mathematics,
Irvine, CA 92697-3875,  USA.
 Email: {aklein@uci.edu}}}

\date{}

\maketitle

\begin{abstract}
We study the region of complete localization in a class of random
 operators which includes  random   Schr\"odinger operators with Anderson-type potentials and classical wave operators in random media,  as well as 
the  Anderson tight-binding model.   We establish new characterizations or criteria   
 for this region of complete
localization, given either by the  decay of eigenfunction 
 correlations or by the decay
 of Fermi projections. (These are necessary and sufficient conditions  for  the random operator to exhibit complete localization  in this energy region.)  Using the first type of characterization we prove
that  in the region of complete localization the random 
operator has   eigenvalues with finite multiplicity.
\end{abstract}


\section{Introduction}

 We study localization in a class of random
 operators which includes 
 random   Schr\"odinger operators with Anderson-type potentials and
  classical wave operators in random media,  as well as 
the  Anderson tight-binding model.  For these operators localization 
 is  obtained either by a multiscale analysis
\cite{FS,FMSS,CKM,vD,Sp,VDK,KLS,Kl,FK1,FK2,
CH1,CH2,FK3,FK4,W1,BCH,KSS,CHT,
GK1,St,Kl4,DSS1,GK3,GK4,KK2,K4}, or,
in certain cases, by the 
fractional moment method \cite{AM,A,ASFH,W2,Kl3,AENSS}. In addition to
 pure point spectrum with exponentially localized eigenfunctions, localization
proved by a either a multiscale analysis or the fractional moment method 
always include  other properties such as dynamical localization
 \cite{A,GDB,ASFH,DS,GK1,AENSS}.

In \cite{GK5} we proved a converse to the multiscale analysis: the region of dynamical localization  coincides with the region where the multiscale analysis
(and the fractional moment method, when applicable) can be performed.
We also gave a large list of characterizations of this region
of  localization, that is, necessary and sufficient conditions  to be satisfied by the random operator in this energy region for a multiscale analysis to be performed at these energies  \cite[Theorem 4.2]{GK5}.
This region of  localization  is the analogue for random operators of the
 region of complete
analyticity for classical spin systems \cite{DS1,DS2}. For this reason we call it the
 \emph{region of complete 
localization}. (Note that the spectral region of complete localization
 is called the strong insulator region
in \cite{GK5}, and the region of complete localization is called the region of
dynamical localization in \cite{GKS}.)

In this article we establish two new consequences of the multiscale analysis
that are also characterizations of  the region of complete
localization,  given either by the  decay of eigenfunction correlations or
by the  decay of Fermi projections. 
Using the characterization by  the decay of eigenfunction correlations we prove
that  in the region of complete localization   the random 
operator has eigenvalues with finite multiplicity.   

 In the one-dimensional case the multiplicity of eigenvalues is easily seen to
be always  less than
 or equal to $2$.  But for $d >1$
 this had only been previously known for in two cases. 
The first is the Anderson tight-binding  model  with bounded density for the probability distribution of the
 single site potential, which has simple eigenvalues in the region  of localization
 \cite{Si,KM}.  The second is its continuum analogue, 
 Anderson-type Hamiltonians in the continuum  with bounded density for the
 probability distribution of the strength of  single site potential, for which
  the  finite multiplicity of eigenvalues in the region of localization is  known \cite{CH1}. 
 (Although Simon's original proof for the Anderson model \cite{Si} does not shed light on the continuum, the recent proof by Klein and Molchanov \cite{KM} indicates that these  Anderson-type Hamiltonians in the continuum should have simple eigenvalues
in the region of localization.  The missing ingredient is a continuous analogue of Minami's estimate \cite{M}.)  
 
 Our proof of finite multiplicity of eigenvalues only requires 
 the conditions for the multiscale analysis, so it applies in great generality.   It neither  requires probability distributions with bounded densities, nor the unique continuation property for eigenfunctions, both requirements for the Combes and Hislop result \cite{CH1}.  In particular, our result applies to random Landau Hamiltonians
 \cite{CH2,W1,GK3,GKS} and to classical wave operators (e.g., acoustic and Maxwell operators) in random media \cite{FK3,FK4,KK2}.

We first characterize the region of complete localization by the decay of the
 expectation of eigenfunction correlations (Theorem~\ref{thmsgenSUDEC}).
 We call this characterization
the strong form of ``Summable Uniform Decay of Eigenfunction Correlations" (SUDEC).
SUDEC has also an almost-sure version which is essentially equivalent 
to the SULE (``Semi Uniformly Localized Eigenfunctions") property introduced in
{\cite{DRJLS0,DRJLS}}.   This almost-sure SUDEC  is a modification
of  the WULE (``Weakly Uniformly Localized Eigenfunctions") property
in \cite{Ge}.  (See also \cite{T} for related properties.)
 But although SUDEC has a strong form (i.e., in expectation),
SULE does not by its very definition.

Recently detailed almost-sure properties of localization like SULE or 
SUDEC, which go beyond  exponential localization or
 almost-sure  dynamical localization, 
turned out to be crucial in the
analysis of the quantum Hall effect. In \cite{EGS}, SULE is used to prove the
equivalence between edge and bulk conductance in quantum Hall systems whenever
the Fermi energy falls into a region of localized states. In \cite{CG,CGH}, SUDEC is
used to regularize the edge conductance in the region of localized states and get
its quantization to the desired value. In \cite{GKS}, SUDEC is the main 
ingredient for
a new and quite transparent proof of the constancy of the bulk conductance
if  the Fermi energy lies in a region of localized states.

It is well known that in the region of complete localization
the random operator 
has pure point spectrum  with exponentially decaying eigenfunctions
\cite{FMSS,VDK,K4}.  The SULE property is also known with 
exponentially decaying eigenfunctions
 \cite{GDB,GK1}.  Theorem~\ref{thmsgenSUDEC} yields easily
an almost-sure SUDEC (and SULE) with sub-exponentially decaying eigenfunctions.
Combining the 
 proof of \cite[Theorem~1.5]{Ge} with the argument in \cite{VDK,K4},
we obtain a form of SUDEC with exponentially decaying eigenfunctions
(Theorem~\ref{thmsudecapp}).  (See \cite{GK6}  for more on SUDEC and SULE.)

We conclude with a  characterization of
 the region of complete localization by the decay of the expectation
of the operator kernel of Fermi projections (Theorem~\ref{thmfermi}), a  crucial ingredient
in linear response theory    and in explanations of the quantum Hall effect 
\cite{BES,AG,BGKS,GKS}.

The derivation of SUDEC and of the decay of Fermi projections from the 
multiscale analysis is based on the 
methods developed in \cite{GK1} and, in the case of the Fermi projections,
the sub-exponential kernel decay for Gevrey-like functions of generalized
Schr\"odinger operators
given in \cite{BGK}.  That they characterize the region of
complete localization relies on the converse to the multiscale analysis,
the fact that slow transport  implies that a multiscale analysis can be performed
\cite{GK5}.

This article is organized as follows:  We introduce random operators,
state our assumptions, and define the region of complete localization
in Section~\ref{secassump}. We state our results in Section~\ref {secthm}.
Theorem~\ref{thmsgenSUDEC} and its corollaries are proved in
 Section~\ref{secproofSUDEC}. Theorem~\ref{thmsudecapp} is proved
in Section~\ref{secexpdec}.  The proof of Theorem~\ref{thmfermi} is given
in Section~\ref{secproofFermi}.

{\sf Notation:}  We set $ \langle x \rangle :   = \sqrt{1+|x|^2} $
for $x \in \R^d$.  By $\Lambda_L(x)$ we denote the open cube (or box)
$\Lambda_L(x)$ in $\R^d$ (or $\Z^d$), centered at $x \in \Z^d$ with side of length
  $L> 0$; we write $\chi_{x,L}$ for its
characteristic function, 
and set  $\chi_{x}:=\chi_{x,1}$. 
Given an open interval $I \subset \mathbb{R}$, we  denote by
$C^\infty_c (I)$  the class
of  real valued infinitely differentiable functions on 
$\mathbb{R}$ with compact
support contained in $I$, with $C^\infty_{c,+} (I)$ being
 the subclass of nonnegative functions. The Hilbert-Schmidt norm of an operator $A$ is
written as
$\|A\|_2$, i.e., $\|A\|_2^2 =\tr  A^*A$.
$C_{a,b, \ldots}$, $K_{a,b, \ldots}$, etc., 
 will always denote some finite constant depending only on 
$a,b, \ldots$.
 (We  omit the dependence on the dimension $d$ in  final results.)

\section{Random operators and the region of complete localization}\label{secassump}

In this article \emph{a random operator is a $\mathbb{Z}^d$-ergodic
 measurable map
 $H_\omega$
from a probability space $(\Omega,\mathcal{F},\mathbb{P})$
(with  expectation $\mathbb{E}$) to generalized 
Schr\"odinger operators on the Hilbert space $\H$, where
 either 
$\H=\mathrm{L}^2(\mathbb{R}^d,{\rm d}x; \mathbb{C}^n)$
or $\H=\ell^2(\mathbb{Z}^d; \mathbb{C}^n)$. } 
Generalized 
Schr\"odinger operators  are a class of
 semibounded second order partial differential
operators of Mathematical Physics, which includes the
Schr\"odinger operator, the  magnetic Schr\"odinger operator,
 and the classical wave operators, eg., the acoustic operator and the
Maxwell operator. (See \cite{GK2} for a precise definition and \cite{K4} for examples.)
 \emph{We assume that $H_\omega$ satisfies the standard conditions
for a generalized Schr\"odinger operator with constants uniform in $\omega$.}

Measurability of $H_\omega$ 
means that
 the mappings $\omega \to f(H_\omega)$ are weakly
(and hence strongly)
measurable for all bounded Borel  measurable functions $f$ on $\mathbb{R}$.
 $H_\omega$ is  $\mathbb{Z}^d$-ergodic if 
there exists  a group representation of  $\mathbb{Z}^d$ by an ergodic family
$\{ \tau_y; \ y \in \mathbb{Z}^d\}$ of measure preserving
 transformations
on $ (\Omega, \mathcal{F}, \mathbb{P})$ 
 such that  we have the covariance given by
\begin{equation} \label{cov}
U(y) H_\omega U(y)^*=  H_{\tau_y(\omega)}\;\; 
\mbox{for  all $ y \in \mathbb{Z}^d$},
\end{equation}
where $U(y)$ is the unitary operator given
by translation: $(U(y)f)(x) = f(x-y)$.  (Note that for Landau Hamiltonians
translations are replaced
by magnetic translations.) It follows 
 that there exists a nonrandom set $\Sigma $
such that $\sigma (H_\omega)=\Sigma $ with probability one, where 
$\sigma (A)$ denotes the spectrum of the operator $A$. In addition, the
decomposition of $\sigma (H_\omega)$ into pure point spectrum,
absolutely continuous spectrum,
 and singular continuous spectrum   is also the same
 with probability one. (E.g., 
 \cite{PF,St}.)

\emph{We  assume that the random operator $H_\omega$
  satisfies the hypotheses of \cite{GK1,GK5} in an open energy interval $\I$.}
 These were called  assumptions or properties
SGEE, SLI, EDI, IAD, NE, and W  in \cite{GK1,GK3,GK5,K4}.  (Although
the results in  \cite{GK5} are written for random Schr\"odinger operators,
 they hold without change for generalized 
Schr\"odinger operators as long as these hypotheses are satisfied.) Although we assume
a polynomial Wegner estimate as in \cite{GK5}, our results are still valid
if we only have a sub-exponential Wegner estimate, with the caveat that one must
substitute sub-exponential moments for polynomial moments (see 
\cite[Remark 2.3]{GK5}).  In particular, our results apply to 
Anderson or Anderson-type Hamiltonians without the requirement of a
bounded density for the
 probability distribution of the single site potential.

Property SGEE guarantees the existence
of   a  generalized eigenfunction expansion in the strong sense.
\emph{We assume that $H_\omega$ satisfies the stronger trace estimate
 \cite[Eq. (2.36)]{GK1}}, 
as in \cite{GK5}. 
(Note that for classical wave operators we always project to the orthogonal
 complement of the kernel of $H_\omega$, see \cite{GK1,KKS,KK2}.)  For some fixed 
$\kappa > \frac d 2$ (which will be generally omitted from the notation) we let
 $T_a$ denote the operator on $\H$ given by multiplication by the function
$ \la x-a\ra^\kappa$,  $a \in \Z^d$, with $T:=T_0$.
Since $\langle a +b  \rangle \le \sqrt{2}\langle a \rangle
\langle b\rangle$, we have
\begin{equation}\label{Tab}
\| T_b T_a^{-1} \| \le 2^{\frac  \kappa  2} \la b-a \ra^\kappa.
\end{equation}
The  domain of $T$,  $\mathcal{D}(T)$, equipped with the norm
$\|\phi\|_+ = \|T\phi\|$, is a Hilbert space,  denoted by
$\mathcal{H}_+\,$.  The Hilbert space $\mathcal{H}_-$ is defined as
 the completion of  $\mathcal{H}$ in the norm $\|\psi\|_- = \|T^{-1}\psi\|$.
 By construction,
$\mathcal{H}_+\subset\mathcal{H}\subset\mathcal{H}_-\,$, and the
natural injections  
 $\imath_+: \mathcal{H}_+\rightarrow\mathcal{H}$  and 
$\imath_-:\mathcal{H}\rightarrow\mathcal{H}_-$ are
continuous with dense range.  The operators $T_+: \mathcal{H}_+ \to \mathcal{H}$
and $T_-: \mathcal{H} \to \mathcal{H}_-$, defined by
$T_+= T \imath_+\,$, and 
 $T_- = \imath_- T $ on $ \mathcal{D}(T)$, are unitary.
We define the random spectral measure
\begin{equation}\label{trestmeas}
\mu_\omega({B}):= \tr \{T^{-1}P_{{B},\omega}T^{-1}\}=\|T^{-1}P_{{B},\omega}\|_2^2,
\end{equation}
where ${B}\subset \R$ is a Borel set and
$P_{{B},\omega} = \chi_{{B}}(H_{\omega})$.  It follows from \cite[Eq. (2.36)]{GK1}
that for $\P$-a.e.  $\omega$ we have 
\begin{equation}\label{trest}
{\mu_\omega({B})}={\mu_\omega({B}\cap \Sigma)} \le  K_{{B}\cap \Sigma},
\end{equation}
where $K_B:= K_{{B}\cap \Sigma}$ is independent of $\omega$, increasing in 
 ${B}\cap \Sigma $, and  $K_B< \infty$ if ${B}\cap \Sigma $ is bounded. Using the  covariance \eqref{cov}, for $\P$-a.e.  $\omega$ and all $a \in \Z^d$
we have
\begin{equation}\label{trest2}
{\mu_{a,\omega}({B})}:= \|T_a^{-1}P_{{B},\omega}\|_2^2= 
\|T^{-1}P_{{B},\tau(- a)\omega}\|_2^2= {\mu_{\tau(- a)\omega}({B})}
\le K_{{B}}.
\end{equation}
We have a generalized eigenfunction expansion for $H_\omega$: For $\P$-a.e. 
$\omega$
 there exists a $\mu_\omega$-locally integrable function 
$\Pb_\omega(\lambda)\colon \R \to {\mathcal{ T}_1(\mathcal{H}_+,\mathcal{H}_-)}$,
the Banach space of bounded operators  $A\colon \mathcal{H}_+ \to
\mathcal{H}_-$  with  $T_-^{-1}AT_+^{-1}$ trace class,
such that
\begin{equation}\label{a16a}
\tr \left\{T_-^{-1}\Pb_\omega(\lambda)T_+^{-1}\right\}=1
\quad  \text{for          
$\mu_\omega$-a.e.\  $\lambda$},
\end{equation}
and, for all Borel sets $B $ with ${B}\cap \Sigma $  bounded,
\begin{equation}\label{geneigexp}
\imath_- P_\omega(B) \imath_+=
\int_{B}\Pb_\omega(\lambda)\,d\mu_\omega(\lambda),
\end{equation} 
where the integral is the Bochner integral of 
${\mathcal{ T}_1(\mathcal{H}_+,\mathcal{H}_-)}$-valued 
functions.  
Moreover, if $\phi\in\mathcal{H}_+$, then 
$\Pb_\omega(\lambda)\phi\in\mathcal{H}_-$ is a generalized
eigenfunction of
$H_\omega$ with generalized eigenvalue $\lambda$ 
 (i.e., an eigenfunction of the closure of $H_\omega$
in $\H_-$ with eigenvalue $\lambda$) for 
$\mu_\omega$-a.e  $\lambda$.  (See \cite[Section 3]{KKS} for details.) 

The multiscale analysis requires the notion of a 
{\em finite volume operator}, a ``restriction"
$H_{\omega,x,L}$ of $H_{\omega}$ to the cube (or box)
$\Lambda_L(x)$, centered at $x \in \Z^d$ with side of length  $L\in 2\N$
(assumed here for convenience; we may take   $L\in L_0\N$ for a suitable $L_0 \ge 1$
 as in \cite{GKS}),  where the ``randomness based
 outside the cube $\Lambda_L(x)$" is not taken into account.  \emph{We assume
the existence of appropriate finite volume operators $H_{\omega,x,L}$
for  $x \in \Z^d$ with  $L\in 2\N$
satisfying properties SLI, EDI, IAD, NE, and W in the open interval $\I$}.
 (See the discussion in \cite[Section~4]{GKS}.)


\emph{The region of complete localization  $\Xi^{\text{CL}}_\I$ for the random operator
$H_\omega$ in the open interval $I$
is defined as the set of energies  $E \in \I$  where  we have the conclusions of the bootstrap 
multiscale analysis}, ie., as the
 set of $E \in\I$ for which there exists some
open interval $I\subset \I$, with  $E\in I $, such that 
 given any $\zeta$, $0<\zeta<1 $, and $\alpha$,
 $ 1<\alpha<\zeta^{-1}$, there is a length scale
$L_0\in 6 \mathbb{N}$
 and a mass $m>0$, so if we
 set 
$L_{k+1} = [L_k^\alpha]_{6\mathbb{N}}$, $k=0,1,\dots$,
 we have
\begin{equation} \label{MSAest}
\mathbb{P}\,\left\{R\left(m, L_k, I,x,y\right) \right\}\ge 1
-\mathrm{e}^{-L_{k}^\zeta}
\end{equation} for all $k=0,1,\ldots$, and  $x, y \in \mathbb{Z}^d$
with 
$|x-y| > L_k +\varrho$, where 
\begin{eqnarray}
\lefteqn{R(m,L, I,x,y) =} \label{defsetR} \\ 
&&\{\mbox{$\omega$; for every}
\; E^\prime \in I
\ \mbox{either} \ \Lambda_L(x) \ \mbox{or} \ 
\Lambda_L(y) \  \mbox{is} \  \mbox{$(\omega,m, E^\prime)$-regular} \} \ .
\nonumber
\end{eqnarray} 
Here $[K]_{6\mathbb{N}}= \max \{ L \in 6\mathbb{N}; \; L \le K\}$
(we work with scales in $6 \N$ for convenience); $\rho>0$ is given in Assumption IAD,
if $\mathrm{dist}(\Lambda_L(x),\Lambda_{L^\prime}(x^\prime) ) 
>\varrho$, then events based in $\Lambda_L(x)$ and
 $\Lambda_{L^\prime}(x^\prime)$ are independent. Given $E \in \mathbb{R}$,
$x \in  \mathbb{Z}^d$ and $L \in 6 \mathbb{N}$,  
we say that
 the box 
$\Lambda_L(x) $ is
$(\omega,m, E)$-regular  for a
 given $m>0$ if $E \notin \sigma(H_{\omega,x,L})$ and 
\begin{equation} 
\| \Gamma_{x,L} R_{\omega,x,L}(E)\chi_{x,{{\frac L 3}}} 
\| \le {\rm e}^{-m\frac{L}{2}} 
 , \label{regular}
\end{equation}
where $R_{\omega,x,L}(E) = (H_{\omega,x,L}-E)^{-1}$ and $\Gamma_{x,L}$
denotes the charateristic function of the ``belt" 
$ \overline{\Lambda}_{L-1}(x) \backslash {\Lambda}_{L-3}(x) $.  (See \cite{GK1,K4}.
 We will take
$\H=\mathrm{L}^2(\mathbb{R}^d,{\rm d}x; \mathbb{C}^n)$,
but the arguments can be easily modified for
 $\H=\ell^2(\mathbb{Z}^d; \mathbb{C}^n)$.)

By construction $\Xi^{\text{CL}}_\I$ is an open set.   It can be defined in many ways,
we gave the most convenient definition for our purposes. 
(We refer to \cite[Theorem~4.2]{GK5} for the equivalent properties
that characterize  $\Xi^{\text{CL}}_\I$.  The spectral region of complete localization in $\I$, 
 $\Xi^{\text{CL}}_\I\cap \Sigma $, is called the ``strong insulator region" in
 \cite{GK5}.)
 Note that
 $\Xi^{\text{CL}}_\I$ is the set of energies in $\I$ where we can perform the bootstrap 
multiscale analysis. 
 (If the conditions
for the fractional moment method are satisfied in $\I$, $\Xi^{\text{CL}}_\I$
 coincides with the set of 
energies  in $\I$ where the fractional moment method can be performed.) 
By our definition spectral gaps are (trivially) intervals of complete
 localization.

\section{Theorems and corollaries}\label{secthm}

 In this article we provide 
two new characterizations of the region of  complete localization.  
The first  characterizes the region of  complete localization by the decay of the
 expectation of generalized eigenfunction correlations, the second by the
expectation of  decay of Fermi projections. 

We start with generalized eigenfunctions.
Given $\lambda \in \R$ and $a \in \Z^d$ we set
\begin{align} \label{defGWx}
\W_{\lambda,\omega}(a):=\begin{cases} 
\displaystyle{\sup_{\substack{\phi \in \H_+\\
\Pb_{\omega}(\lambda)\phi  \not= 0}}} 
\ \frac {\| \chi_a \Pb_{\omega}(\lambda)\phi \|}
{\|T_a^{-1}\Pb_{\omega}(\lambda)\phi \|}&
 \text{if $\Pb_{\omega}(\lambda)\not=0$},\\0 & \text{otherwise},\end{cases}
\end{align}
$\W_{\lambda,\omega}(a)$ is a
measurable function of  $(\lambda, \omega)$ for each $a\in \Z^d$ with
\begin{equation}\label{boundGW}
\W_{\lambda,\omega}(a)\le 
\scal{\tfrac {\sqrt{d}} 2}^\kappa= \left(1 + \tfrac d 4\right)^{\frac \kappa 2} .
\end{equation}

Our first characterization is given in the following theorem.

\begin{theorem}\label{thmsgenSUDEC}  Let $I$ be a bounded open interval
with  $\bar{I} \subset \I$.
If $\bar{I} \subset \Xi^{\text{CL}}_\I$, then  
 for all $\zeta\in]0,1[$ we have
\begin{equation}\label{sgenSUDEC}
\E \left\{ \norm{ \W_{\lambda,\omega}(x)
 \W_{\lambda,\omega}(y)}_{\L^\infty(I,\di \mu_\omega(\lambda))}\right\} \le
 C_{I,\zeta}  
\,  \e^{-|x-y|^\zeta}  \quad  \text{for  all $ x,y \in \Z^d$}.
\end{equation}
Conversely, if  \eqref{sgenSUDEC}  holds for some  $\zeta\in]0,1[$,
then $I \subset \Xi^{\text{CL}}_\I$.
\end{theorem}

Note that the converse will still hold if we only have fast enough polynomial decay
in \eqref{sgenSUDEC}.

\begin{remark}  We may replace the
denominator $\|T_a^{-1}P_{\lambda,\omega}\phi \|$
in \eqref{defGWx} by $$\Theta_a (\phi) := \inf_{b
\in \Z^2} 
 \left\{ \la b-a\ra^\kappa \norm{T_b^{-1}P_{\lambda,\omega}\phi}\right\}.$$
Since $\Theta_a (\phi)  \le 
 \norm{T_a^{-1}P_{\lambda,\omega}\phi}$,
 this slightly improves \eqref{sgenSUDEC}.
\end{remark}

\begin{corollary}\label{corpurepoint}
$H_\omega$ has pure point spectrum in the open set
  $\Xi^{\text{CL}}_\I$
 for $\P$-a.e.  $\omega$,  with the corresponding eigenfunctions decaying faster than any
sub-exponential.  Moreover,  we have 
 (with $P_{\lambda,\omega}:=P_{\{\lambda\},\omega} $)
\begin{align}\label{decayWP}
\E \left\{\norm{ \mu_\omega(\{\lambda\})
\left(\tr  P_{\lambda,\omega}\right) 
}_{\L^\infty(I,\di \mu_\omega(\lambda))}\right\} \le C_{I}< \infty,
\end{align}
 and hence 
 for $\P$-a.e.  $\omega$ the eigenvalues of  $H_\omega$ in $\Xi^{\text{CL}}_\I$
are of finite multiplicity.
\end{corollary}

It is well known that  $H_\omega$ has pure point spectrum in
$\Xi^{\text{CL}}_\I$ with exponentially decaying eigenfunctions.  
Our point is that pure point spectrum follows directly from \eqref{sgenSUDEC},
also yielding sub-exponentially decaying eigenfunctions. The estimate
\eqref{decayWP} is new, and it immediately implies that
 for $\P$-a.e.\ $\omega$ 
 the random operator  $H_\omega$ has only
eigenvalues with finite multiplicity in $\Xi^{\text{CL}}_\I$.

 If $H_\omega$ has pure point spectrum we might as well  work with 
eigenfunctions, not generalized eigenfunctions. 
Given $\lambda \in \R$ and $a \in \Z^d$ we set
\begin{align} \label{defWx}
W_{\lambda,\omega}(a):=\begin{cases} 
\displaystyle{\sup_{\substack{\phi \in \H\\
P_{\lambda,\omega}\phi  \not= 0}}} 
\ \frac {\| \chi_a P_{\lambda,\omega}\phi \|}
{\|T_a^{-1}P_{\lambda,\omega}\phi \|}&
 \text{if $P_{\lambda,\omega}\not=0$},\\0 & \text{otherwise},\end{cases}
\end{align}
and
\begin{align} \label{defZx}
Z_{\lambda,\omega}(a):=\begin{cases} 
\displaystyle{\frac {\| \chi_a P_{\lambda,\omega} \|_2}
{\|T_a^{-1}P_{\lambda,\omega}\|_2}} &
 \text{if $P_{\lambda,\omega}\not=0$},\\0 & \text{otherwise}.\end{cases}
\end{align}
$W_{\lambda,\omega}(a)$ and $Z_{\lambda,\omega}(a)$ are
measurable functions of  $(\lambda, \omega)$ for each $a\in \Z^d$.
They are 
  covariant, that is,
\begin{equation}\label{Wcov}
Y_{\lambda,\omega}(a) = Y_{\lambda,\tau(b) \omega}(a+b )
\quad \text{for all $b \in \Z^d$, with $Y=W$ or $Y=Z$ }.
\end{equation}

 It follows from 
\eqref{geneigexp} that 
$\imath_- P_{\lambda,\omega} \imath_+=
\Pb_{\omega}(\lambda)\mu_\omega(\{\lambda\})$.   Since 
$P_{\lambda,\omega}\not=0$ if and only if  $\mu_\omega(\{\lambda\})\not=0$,
we  have  $W_{\lambda,\omega}(a)= \W_{\lambda,\omega}(a)$ if 
$\mu_\omega(\{\lambda\})\not=0$ and  $W_{\lambda,\omega}(a)=0$ otherwise.
Combining this fact with 
the definition of the Hilbert-Schmidt norm and \eqref{boundGW} we get
\begin{equation}\label{boundW}
Z_{\lambda,\omega}(a) \le W_{\lambda,\omega}(a)\le  \W_{\lambda,\omega}(a)\le
 \left(1 + \tfrac d 4\right)^{\frac \kappa 2} .
\end{equation}

\begin{remark} \label{rempurepoint} $H_\omega$ has pure point spectrum in an open interval $I$
if and only if for
$\P$-a.e.\ $\omega$ we have
 $W_{\lambda,\omega}(a)=\W_{\lambda,\omega}(a)$ for
all $a \in \Z^d$ and $\mu_\omega$-a.e. 
$\lambda \in  I$. 
\end{remark}

 Thus we have the following  corollary to
Theorem~\ref{thmsgenSUDEC}.

\begin{corollary}  
\label{corsSUDEC}
Let $I$ be a bounded open interval
with  $\bar{I} \subset \I$.
If $\bar{I} \subset \Xi^{\text{CL}}_\I$,  $H_\omega$ has pure point spectrum in 
$\bar{I} $ for $\P$-a.e. $\omega$ and
 for all  $\zeta\in]0,1[$ and $x,y \in \Z^d$ we have
\begin{align}\label{decayZ}
\E \left\{ \norm{ Z_{\lambda,\omega}(x)
 Z_{\lambda,\omega}(y)}_{\L^\infty(I,\di \mu_\omega(\lambda))}\right\}& \le 
 C_{I,\zeta} \,\e^{-|x-y|^\zeta}  ,\\
\E \left\{ \norm{ W_{\lambda,\omega}(x)
 W_{\lambda,\omega}(y)}_{\L^\infty(I,\di \mu_\omega(\lambda))}\right\}& \le
 C_{I,\zeta}  \, \e^{-|x-y|^\zeta}  .\label{decayW}
\end{align}
Conversely, if $H_\omega$ has pure point spectrum in 
${I} $ for $\P$-a.e. $\omega$, and either \eqref{decayZ} or \eqref{decayW}
holds for some $\zeta\in]0,1[$, we have  ${I} \subset \Xi^{\text{CL}}_\I$.
\end{corollary}

We now turn to almost sure consequences of Theorem~\ref{thmsgenSUDEC}.

\begin{corollary}\label{thmae}
 Let $I $ be
be a bounded open interval
with  $\bar{I} \subset \Xi^{\text{CL}}_\I$. The following holds
for $\P$-a.e. $\omega$: $H_\omega$ has pure point spectrum in $I$ with finite
multiplicity, so  let  $\{E_{n,\omega}\}_{n \in \N}$ be an enumeration of  the 
(distinct) eigenvalues
 of   $H_{\omega}$ in $I$, with  $\nu_{n,\omega} $ being the (finite) multiplicity
of the eigenvalue  $E_{n,\omega}$.
Then: \\
\textbf{(i)} Summable Uniform Decay of Eigenfunction Correlations (SUDEC):
  For each  $\zeta \in ]0,1[$ and $\eps > 0$
 we have 
\begin{align}
\lVert \chi_x\phi \rVert\lVert \chi_y\psi  \rVert &\le
 C_{I,\zeta,\eps,\omega}  \|T^{-1}\phi\|\|T^{-1}\psi\|
 \la  y\ra^{d +\eps}
\  \e^{-|x-y|^\zeta} \label{decayphi2222},\\
\lVert \chi_x\phi \rVert\lVert \chi_y\psi  \rVert  &\le
 C_{I,\zeta,\eps,\omega}   \|T^{-1}\phi\|\|T^{-1}\psi\|
 \la  x\ra^{\frac {d +\eps }2} \la  y\ra^{\frac {d +\eps } 2}
\  \e^{-|x-y|^\zeta} \label{decayphi22},
\end{align}
for all  $\phi,\psi \in  \Ran P_{E_{n,\omega},\omega}$,  $n \in \N$,  and $ x,y \in \Z^d$. 
\\
\textbf{(ii)} Semi Uniformly Localized Eigenfunctions (SULE): There exist centers of localization $\{y_{n,\omega}\}_{n\in \mathbb{N}}$
for the eigenfunctions
such that
for each  $\zeta \in ]0,1[$ and $\eps >0$  we have 
\begin{align}
\lVert \chi_x {\phi }  \rVert 
 \le    C_{I,\zeta,\eps,\omega}   \|T^{-1}\phi\|
 \la y_{n,\omega}\ra^{2(d+\eps)}
\  \e^{- |x-y_{n,\omega}|^\zeta},   \label{SULE}  
\end{align}
for all $\phi  \in  \Ran P_{E_{n,\omega},\omega}$, $n \in \N$,   and $ x \in \Z^d$. Moreover, we have
\begin{equation}\label{NL}
N_{L,\omega} :=\sum_{n \in \N; |y_{n,\omega}| \le L } \nu_{n,\omega}  \le 
 C_{I,\omega} L^{d }\quad \text{for all} 
\   L \ge 1.
\end{equation}
\textbf{(iii)} SUDEC and SULE for complete orthonormal sets:  For each $n \in \N$  let
$\{\phi_{n,j,\omega}\}_{ j\in \{1,2,\dots,\nu_{n,\omega}\} }$
be an orthonormal basis for the eigenspace  $\Ran P_{E_{n,\omega},\omega}$,
so $\{\phi_{n,j,\omega}\}_{ n\in \N, j\in \{1,2,\dots,\nu_{n,\omega}\} }$
is a  complete orthonormal set  of eigenfunctions of $H_\omega$ with energy in $I$.
 Then  for each  $\zeta \in ]0,1[$ and $\eps > 0$
 we have 
\begin{align}
\lVert \chi_x\phi_{n,i,\omega}  \rVert\lVert \chi_y\phi_{n,j,\omega}  
\rVert &\le
 C_{I,\zeta,\eps,\omega} \sqrt{\alpha_{n,i,\omega}}\sqrt{\alpha_{n,j,\omega}}
 \la  y\ra^{d +\eps}
\  \e^{-|x-y|^\zeta} \label{decayphi22225},\\
\lVert \chi_x\phi_{n,i,\omega}  \rVert\lVert \chi_y\phi_{n,j,\omega}  
\rVert &\le
 C_{I,\zeta,\eps,\omega}  \sqrt{\alpha_{n,i,\omega}}\sqrt{\alpha_{n,j,\omega}}
 \la  x\ra^{\frac {d +\eps }2} \la  y\ra^{\frac {d +\eps } 2}
\  \e^{-|x-y|^\zeta} \label{decayphi225},\\
\lVert \chi_x {\phi_{n,j,\omega} }  \rVert 
& \le    C_{I,\zeta,\eps,\omega} \sqrt{\alpha_{n,j,\omega}}
 \la y_{n,\omega}\ra^{2(d+\eps)}
\  \e^{- |x-y_{n,\omega}|^\zeta},   \label{SULE5}  
\end{align}
for all $n \in \N$, $i,j \in \{1,2,\dots,\nu_{n,\omega}\}$,  and $ x,y \in \Z^d$,
where  
\begin{gather}\label{defalphan}
\alpha_{n,j,\omega} :=  \|T^{-1}\phi_{n,j,\omega}\|^2 , \quad \text{$n \in\N$, 
$ j\in \{1,2,\dots,\nu_{n,\omega}\}$},\\
\label{sumalphan}
 \sum_{ j\in \{1,2,\dots,\nu_{n,\omega}\}} 
 \alpha_{n,j, \omega}= \mu_\omega({\{E_{n,\omega}\}})  \quad \text{for all $n \in \N$},\\
\label{sumalpha}
\sum_{n, \in \N,  j\in \{1,2,\dots,\nu_{n,\omega}\}} 
 \alpha_{n,j, \omega}= \sum_{n \in \N} \mu_\omega({\{E_{n,\omega}\}})= 
\mu_\omega(I).
\end{gather}
\end{corollary}

\begin{remark} The statements (i) and (ii) are essentially equivalent, 
 and  imply finite  multiplicity for eigenvalues,  while (iii) does not, see \cite{GK6}.
Note that in (ii)  eigenfunctions associated to the same eigenvalue  have
the same center of localization. It is easy to see that \eqref{decayphi2222} implies 
\eqref{decayphi22}, the reverse implication also being true up to a change in the
constant--both forms of SUDEC are useful.
\end{remark}

 If $I $ is a bounded open interval
with  $\bar{I} \subset \Xi^{\text{CL}}_\I$, it is known that that 
for $\P$-a.e. $\omega$ the operator $H_\omega$
has pure point spectrum in $I$ with exponentially decaying eigenfunctions
\cite{FMSS,VDK,K4}.
 The SULE property is also known with exponential decay \cite{GDB,GK1}.
Combining the 
 proof of \cite[Theorem~1.5]{Ge} with the argument in \cite{VDK,K4}
we also obtain SUDEC with exponential decay for  $\P$-a.e. $\omega$.

\begin{theorem}\label{thmsudecapp}
 Let $I $ be
be a bounded open interval
with  $\bar{I} \subset \Xi^{\text{CL}}_\I$. For all $\phi\in\H_+$ and $\lambda\in I$
  set $\alpha_{\lambda,\phi} :=  \|T^{-1} \Pb_\omega(\lambda) \phi\|^2$.
 The following holds
for $\P$-a.e. $\omega$ and $\mu_\omega$-a.e. $\lambda \in I$: 
 For all $\eps > 0$  there exists $m_\eps=m_{I,\eps} >0$  such that for all
$\phi,\psi\in\H_+$ we have 
\begin{align} \label{sudecappbis}
& \lVert \chi_x \Pb_\omega(\lambda) \phi  \rVert 
\lVert \chi_y \Pb_\omega(\lambda) \psi
\rVert   \\
& \qquad \qquad  \le 
 C_{I,\eps,\omega} \sqrt{\alpha_{\lambda,\phi}\alpha_{\lambda,\psi}}\,
 \mathrm{e}^{(\log {\la x\ra} )^{1+\eps}}\mathrm{e}^{(\log {\la y\ra} )^{1+\eps}}
 \e^{-m_\eps |x-y|}\nonumber
\end{align}
for all $ x,y \in \Z^d$. In particular, it follows that 
$H_\omega$ has pure point spectrum in $I$ with exponentially decaying 
eigenfunctions.
\end{theorem}

Unlike Theorem~\ref{thmsgenSUDEC}, Theorem~\ref{thmsudecapp} does not give
a characterization of the region of complete localization.
But it still implies that $H_\omega$
 has only eigenvalues with finite multiplicity in $I$ \cite{GK6}. 
 
Compared to the rather short and transparent proof of \eqref{decayphi22}, 
the proof of \eqref{sudecappbis} is quite technical and involved--an extra 
motivation for deriving \eqref{decayphi22}.

We now turn to the characterization in terms of the decay of Fermi projections. 
We set   $P_{\omega}^{(E)}:=P_{]-\infty,E],\omega}$,
the Fermi projection corresponding to the Fermi energy $E$.

\begin{theorem}\label{thmfermi}  
Let $I $ and $I_1$ be
 bounded open intervals with $\bar{I} \subset I_1\subset \bar{I_1} \subset  
\Xi^{\text{CL}}_\I$.
If $\bar{I} \subset \Xi^{\text{CL}}_\I$
Let $I $ be
be a bounded open interval
with  $\bar{I} \subset \I$.
If $\bar{I} \subset \Xi^{\text{CL}}_\I$, then  
 for all  $\zeta\in]0,1[$ we have
\begin{equation} \label{fermidecay}
\mathbb{E}\left\{\sup_{E \in I} \left\|\chi_{x}
P_{\omega}^{(E)}
\chi_{y}\right\|_2^2\right\}
 \leq   C_{I,\zeta} \,\mathrm{e}^{-|x-y|^\zeta} \quad \text{for all $x,y \in \Z^d$}.
\end{equation}  
Conversely, if \eqref{fermidecay}  holds for some   $\zeta\in]0,1[$,
then $I \subset \Xi^{\text{CL}}_\I$.
\end{theorem}

Again,the converse will still hold if we only have fast enough polynomial decay
in \eqref{fermidecay}.  Its proof explicitly uses that slow enough transport 
(weaker than dynamical localization) implies that a multiscale analysis can 
be performed. The estimate \eqref{fermidecay} is known to hold for the 
Anderson model on the lattice with exponential decay, using the 
estimate given by the fractional moment method \cite{AG}.


\section{Summable Uniform Decay of Eigenfunction Correlations}
\label{secproofSUDEC}

In this section we prove Theorem~\ref{thmsgenSUDEC} and its corollaries.

\begin{proof}[Proof of  Theorem~\ref{thmsgenSUDEC}]
Since $\bar{I} \subset \Xi^{\text{CL}}_\I$, 
given any $\zeta$, $0<\zeta<1 $, and $\alpha$,
 $ 1<\alpha<\zeta^{-1}$, there is a length scale
$L_0\in 6 \mathbb{N}$
 and a mass $m>0$, so if we
 set 
$L_{k+1} = [L_k^\alpha]_{6\mathbb{N}}$, $k=0,1,\dots$,
 we have \eqref{MSAest}
for all $k=0,1,\ldots$, and  $x, y \in \mathbb{Z}^d$
with 
$|x-y| > L_k +\varrho$.

Let  $I \subset \Xi^{\text{CL}}_\I$ be a bounded interval with  $\bar{I} \subset \I$.
Note that the quantity
$\norm{ \W_{\lambda,\omega}(x)
\W_{\lambda,\omega}(y)}_{\L^\infty(I,\di \mu_\omega(\lambda))}$ 
is measurable in $\omega$ since the $L^\infty$ norm on  sets of finite measure
is the limit of the $L^p$ norms as $p \to \infty$. (It  is actually covariant in view
of the way 
$\Pb_\omega(\lambda)$ is constructed (see \cite[Eq. (46)]{KKS}), and the fact 
that  the measures $ \mu_\omega$ and
 $ \mu_{\tau(a)\omega}$ are equivalent.)

\begin{lemma} Let  $\omega\in R(m,L, I,x,y) $ (defined in\eqref{defsetR}).
Then
\begin{equation}\label{WWest}
\norm{ \W_{\lambda,\omega}(x)
 \W_{\lambda,\omega}(y)}_{\L^\infty(I,\di \mu_\omega(\lambda))} \le 
C_{I,m}\mathrm{e}^{-m \frac L 4}.
\end{equation}
\end{lemma}

\begin{proof}
Let  $\omega\in R(m,L, I,x,y) $.  Then for any $\lambda\in I$, either 
$\Lambda_L(x)$
or $\Lambda_L(y)$ is $(m,\lambda)$-regular for $H_\omega$,
  say $\Lambda_L(x)$. Given
$\phi\in\mathcal{H}_+$,  $\Pb_{\omega}(\lambda)\phi$ is a generalized eigenfunction
of
$H_\omega$ with  eigenvalue $\lambda$
(perhaps the trivial eigenfunction $0$), so it follows from the EDI
\cite[(2.15)]{GK1}, using $\chi_x=  \chi_{x, \frac L 3}\, \chi_x$, that
\begin{equation}
\| \chi_x\Pb_{\omega}(\lambda)\phi \| \le  
 \tilde{\gamma}_{I} 
\| \Gamma_{x,L} R_{x,L}(\lambda) \chi_{x,L/3}\|_{x,L} 
\|  \Gamma_{x,L} \Pb_{\omega}(\lambda)\phi\| .\label{eigdec2}
\end{equation}
 Since 
 $\Lambda_L(x)$ is $(m,\lambda)$-regular, we have that
\begin{align}
\| \chi_x \Pb_{\omega}(\lambda)\phi \|  \le
 \tilde{\gamma}_{I} 
\mathrm{e}^{-m \frac L 2} \|  \Gamma_{x,L} \Pb_{\omega}(\lambda)\phi\| \le
C_{I,m,d}^\prime \mathrm{e}^{-m \frac L 4}
\| T_x^{-1}\Pb_{\omega}(\lambda)\phi\|,\label{eigdec9}
\end{align}
since
\begin{equation}
 \|\Gamma_{x,L} \Pb_{\omega}(\lambda)\phi\|   \le C_{d}
 L^{d-1}\la \textstyle{ \frac {L+1}  {2}} \ra^{\kappa} \| T_x^{-1}
 \Pb_{\omega}(\lambda)\phi\| .
\end{equation}
Thus, using the bound \eqref{boundGW} for the term in $y$, we get \eqref{WWest}.
\end{proof}

If  $\bar{I} \subset \Xi^{\text{CL}}_\I$, 
given any $\zeta$, $0<\zeta<1 $, and $\alpha$,
 $ 1<\alpha<\zeta^{-1}$, there is a length scale
$L_0\in 6 \mathbb{N}$
 and a mass $m>0$, so if we
 set 
$L_{k+1} = [L_k^\alpha]_{6\mathbb{N}}$, $k=0,1,\dots$,
 we have \eqref{MSAest}
for all $k=0,1,\ldots$, and  $x, y \in \mathbb{Z}^d$
with 
$|x-y| > L_k +\varrho$.

Thus given  $x,y \in \mathbb{Z}^d$  and $k$ such that
 $L_{k+1} +\varrho \geq |x-y|> L_k +\varrho$, it follows from \eqref{WWest} that 
\begin{align}
\mathbb{E}\left\{ \norm{ \W_{\lambda,\omega}(x)
 \W_{\lambda,\omega}(y)}_{\L^\infty(I,\di \mu_\omega(\lambda))} ; 
{R(m,L_k, I,x,y)} \right\}\le C_{I,m }
\mathrm{e}^{- m\frac {L_k}4}.
\end{align}
On the complementary set we use the bound  \eqref{boundGW} for both terms,
obtaining
\begin{align}
&\mathbb{E}\left\{ \norm{ \W_{\lambda,\omega}(x)
 \W_{\lambda,\omega}(y)}_{\L^\infty(I,\di \mu_\omega(\lambda))} ; \omega \notin
{R(m,L_k, I,x,y)} \right\}\\ \notag
& \qquad \qquad  \qquad  \qquad \qquad \le C_{d}\,\P \{\omega \notin
{R(m,L_k, I,x,y)}  \}\le
 C _d  \,\mathrm{e}^{-L_k^\zeta}.
\end{align}
 Since   $L_{k+1} +\varrho \geq |x-y|> L_k +
\varrho$, the estimate \eqref{sgenSUDEC} now follows with
$\frac \zeta \alpha$ instead of $\zeta$.  Since  $\zeta \in ]0,1[ $ and 
$ 1<\alpha<\zeta^{-1}$ are otherwise arbitrary,   \eqref{sgenSUDEC} holds
with any  $\zeta \in ]0,1[ $.

To prove the converse, we use the following lemma.

\begin{lemma} \label{lemxPby} For $\P$-a.e. $\omega$ we have
\begin{equation}\label{xPby}
\norm{\chi_x  \Pb_{\omega}(\lambda) \chi_y}_2^2 \le
   C_d  \la x\ra^{2\kappa}  \la y\ra^{2\kappa}\W_{\lambda,\omega}(x)
 \W_{\lambda,\omega}(y)
\end{equation}
for all $x,y \in \Z^d$, $\lambda \in \R$.
\end{lemma}

\begin{proof}  Let $\{ \psi_n\}_{n \in \N}$ be an orthonormal basis for $\H$.
  We have
\begin{align}\notag
& \norm{\chi_x  \Pb_{\omega}(\lambda) \chi_y}_2^2 = \sum_{n \in \N}
\norm{\chi_x  \Pb_{\omega}(\lambda) \chi_y \psi_n }^2 \\
&\qquad  \le 
 [\W_{\lambda,\omega}(x)]^2  \sum_{n \in \N}
\norm{T_x^{-1}  \Pb_{\omega}(\lambda) \chi_y \psi_n }^2\\
&\qquad  = [ \W_{\lambda,\omega}(x)]^2
\norm{T_x^{-1}  \Pb_{\omega}(\lambda) \chi_y  }_2^2
\le C_d 
\la x\ra^{2\kappa}  \la y\ra^{2\kappa}[\W_{\lambda,\omega}(x)]^2, \notag
\end{align}
where we used \eqref{a16a} and \eqref{Tab}.

Since $\norm{\chi_x  \Pb_{\omega}(\lambda) \chi_y}_2
=\norm{\chi_y  \Pb_{\omega}(\lambda) \chi_x}_2$, the lemma follows.
\end{proof}

So now  assume \eqref{sgenSUDEC} holds for some 
 $\zeta \in ]0,1[ $.  By $\mathcal{B}_1=\mathcal{B}_1(\mathbb{R})$ we denote the
collection of real-valued Borel functions $f$ of a real variable with 
$\sup_{t \in \mathbb{R}}|f(t)| \le 1$. Using the generalized eigenfunction expansion
\eqref{geneigexp},  Lemma~\ref{lemxPby}, and \eqref{trest}, we get 
\begin{align}
&\sup_{f \in\mathcal{B}_1 } 
\left\|\chi_{x}f(H_\omega) P_\omega(I)
\chi_{0}\right\|_2  
\leq 
\sup_{f \in\mathcal{B}_1 }
\int_{I} |f(\lambda)| \left\|\chi_{x} 
\Pb_\omega(\lambda)\chi_{0} \right\|_2
{\rm d}\mu_\omega(\lambda) \label{ineqop} \\
& \ \leq 
\int_{I} \left\|\chi_{x} 
\Pb_\omega(\lambda)\chi_{0} \right\|_2
{\rm d}\mu_\omega(\lambda)
\le C_d ^{\frac 1 2}\la x\ra^{\kappa}  K_I \norm{ \W_{\lambda,\omega}(x)
 \W_{\lambda,\omega}(0)}_{\L^\infty(I,\di \mu_\omega(\lambda))}^{\frac 1 2} .
\notag
\end{align}
Thus it follows from \eqref{sgenSUDEC}  that
\begin{equation}
\E \left\{\sup_{f \in\mathcal{B}_1 } 
\left\|\chi_{x}f(H_\omega) P_\omega(I)
\chi_{0}\right\|_2^2\right\}\le C_d C_{I,\zeta}  K_I^2  \la x\ra^{2\kappa} 
\,  \e^{-|x|^\zeta} \le C^\prime_{I,\zeta}  \,  \e^{-\frac 1 2|x|^\zeta} ,
\end{equation}
and hence for all $x,y \in \Z^d$ we have
\begin{align}\notag
\E \left\{\sup_{f \in\mathcal{B}_1 } 
\left\|\chi_{x}f(H_\omega) P_\omega(I)
\chi_{y}\right\|_2^2\right\}&= \E \left\{\sup_{f \in\mathcal{B}_1 } 
\left\|\chi_{x-y}f(H_\omega) P_\omega(I)
\chi_{0}\right\|_2^2\right\}\\
& \le C^\prime_{I,\zeta}  \,  \e^{-\frac 1 2|x-y|^\zeta} .\label{proofDL}
\end{align}
It now follows from \cite[Theorem 4.2]{GK5} that $I \subset \Xi^{\text{CL}}_\I$
\end{proof}

\begin{proof}[Proof of Corollary~\ref{corpurepoint}]
Let us consider a  bounded interval $I$ with  $\bar{I} \subset \Xi^{\text{CL}}_\I$.
It follows from \eqref{sgenSUDEC89} that
 for any $\phi \in \H_+$ and $\mu_\omega$-a.e. $\lambda \in I$ we have
\begin{align}\notag
\| \chi_x \Pb_{\omega}(\lambda)\phi \| \| \chi_y \Pb_{\omega}(\lambda)\phi \|
&\le2^\kappa  C_{I,\xi,\omega} \e^{- |x-y|^\xi}\la x\ra^{3\kappa}\la y\ra^{\kappa}
\|\phi\|_+^2\\
&\le C_{I,\xi,d,\omega} \la x\ra^{3\kappa}\e^{-\frac 12  |x-y|^\xi}\|\phi\|_+^2
\end{align}
for all $x,y \in \Z^d$,
where we used a consequence of  \eqref{Tab}, namely
\begin{equation}
{\|T_a^{-1}\Pb_{\omega}(\lambda)\phi \|}\le 2^{\frac \kappa 2}\la a\ra^{\kappa}
{\|\Pb_{\omega}(\lambda)\phi \|}_-\le  2^{\frac \kappa 2}\la a\ra^{\kappa}
\|\phi\|_+ \, .
\end{equation}
In particular, if $\Pb_{\omega}(\lambda)\phi\not = 0$ we can pick $x_0 \in \Z^d$
such that $\chi_{x_0} \Pb_{\omega}(\lambda)\phi\not=0$, and thus
\begin{equation}\label{subexpeigdecay}
 \| \chi_y \Pb_{\omega}(\lambda)\phi \| \le C_{I,\xi,d,\omega} 
\| \chi_{x_0}\Pb_{\omega}(\lambda)\phi \|^{-1} \|\phi\|_+^2\la x_0\ra^{3\kappa}
\e^{-\frac 12  |y-x_0|^\xi} \ \text{for all $y \in \Z^d$}.
\end{equation}
It follows  that $ \Pb_{\omega}(\lambda)\phi \in \H$, and  hence
$\mu_\omega$-a.e. $\lambda \in I$ is an eigenvalue of $H_\omega$. Thus
$H_\omega$ has pure point spectrum in $I$, with the corresponding
eigenfunctions decaying faster than any sub-exponential by \eqref{subexpeigdecay}.
(See, e.g., \cite{KKS}.)  

In fact, these eigenvalues have finite multiplicity, a consequence of the estimate
 \eqref{decayWP}, which is proved as follows:
Using \eqref{trest2} and \eqref{boundW}, we have
\begin{equation}
\begin{split}
\mu_\omega(\{\lambda\})\left(\tr  P_{\lambda,\omega}\right)& = \norm{T^{-1}P_{\lambda,\omega}}_2^2
\left(\tr  P_{\lambda,\omega}\right) \\&  \le
C_d \sum_{x,y \in \Z^d}\scal{x}^{-2\kappa}
 \norm {\chi_x P_{\lambda,\omega}}_2^2
 \norm {\chi_y P_{\lambda,\omega}}_2^2\\
& \le C_d K_{I}^2
\sum_{x,y \in \Z^d}\scal{x}^{-2\kappa}
\left( Z_{\lambda,\omega}(x) Z_{\lambda,\omega}(y)\right)^2\\
& \le C^\prime_d K_{I}^2
\sum_{x,y \in \Z^d}\scal{x}^{-2\kappa}
Z_{\lambda,\omega}(x) Z_{\lambda,\omega}(y), \label{boundTP}
\end{split}
\end{equation}
and hence \eqref{decayWP}
 follows from Remark~\ref{rempurepoint} and \eqref{boundW} (or from \eqref
{decayZ}).

\end{proof}

\begin{lemma}\label{lemmaWWp}
Let  $I $ be a bounded interval with  $\bar{I} \subset \Xi^{\text{CL}}_\I$.
Then for all  $\xi \in ]0,1[$,  $p\ge 1$, and  $\P$-a.e. $\omega$ we have
\begin{equation}\label{sgenSUDEC89}
\norm{ \sum_{x,y \in \Z^d}\e^{ |x-y|^\xi}\la x\ra^{-2\kappa} 
\left[\W_{\lambda,\omega}(x)
 \W_{\lambda,\omega}(y)\right]^p}_{\L^\infty(I,\di \mu_\omega(\lambda))}
\hspace{-.4in}\le C_{I,\xi,p,\omega} < \infty.
\end{equation}
\end{lemma}

\begin{proof}
It follows from \eqref{sgenSUDEC} and \eqref{boundGW} that for any $\xi \in ]0,1[$
and $p \ge 1$ we have 
\begin{equation}\label{sgenSUDECyy}
\E \left\{ \sum_{x,y \in \Z^d}\e^{ |x-y|^\xi}\la x\ra^{-2\kappa}\norm{ \W_{\lambda,\omega}(x)
 \W_{\lambda,\omega}(y)}^p_{\L^\infty(I,\di \mu_\omega(\lambda))}\right\} 
 \le C_{I,\xi,p} < \infty,
\end{equation}
and hence \eqref{sgenSUDEC89} follows.
\end{proof}

In fact Lemma~\ref{lemmaWWp} holds for any $p>0$ by modifying the proof of 
Theorem~\ref{thmsgenSUDEC}.

\begin{proof}[Proof of Corollary~\ref{corsSUDEC}]
Since  when $H_\omega$ has pure point spectrum
in ${I} $ for $\P$-a.e. $\omega$ the estimate
 \eqref{decayW} is the same as \eqref{sgenSUDEC},  the corollary with
\eqref{decayW}
follows immediately from Theorem~\ref{thmsgenSUDEC}. The estimate
\eqref{decayZ} follows immediately from  from \eqref{decayW} in view of 
\eqref{boundW}.  To prove the converse from \eqref{decayZ},
note 
that if $\mu_\omega(\{\lambda\}) \not=0$, we have, using \eqref{Tab} and 
\eqref{a16a},
\begin{equation}\label{xPby2}
\begin{split}
& \norm{\chi_x  \Pb_{\omega}(\lambda) \chi_y}_1 =\mu_\omega(\{\lambda\})^{-1}
\norm{\chi_x  P_{\lambda,\omega}\chi_y}_1\\
&\quad  \le \mu_\omega(\{\lambda\})^{-1}
\norm{\chi_x  P_{\lambda,\omega}}_2\norm{\chi_y P_{\lambda,\omega}}_2\\
&\quad  = \mu_\omega(\{\lambda\})^{-1}\norm{T_x^{-1}  P_{\lambda,\omega}}_2
\norm{T_y^{-1}  P_{\lambda,\omega}}_2
Z_{\lambda,\omega}(x)Z_{\lambda,\omega}(y)\\
&  \quad
\le
   C^\prime_d  \la x\ra^{\kappa}  \la y\ra^{\kappa}
 Z_{\lambda,\omega}(x)Z_{\lambda,\omega}(y).
\end{split}\end{equation}
Thus, if $H_\omega$ has pure point spectrum in $I$,
\eqref{proofDL} follows from \eqref{decayZ}, and hence
$I \subset \Xi^{\text{CL}}_\I$ by \cite[Theorem 4.2]{GK5}.
\end{proof}

\begin{proof}[Proof of Corollary~\ref{thmae}]  Pure point spectrum
almost surely in $I$ with eigenvalues of finite multiplicity follows from
 Corollary~\ref{corpurepoint}.
It follows from Lemma~\ref{lemmaWWp} that for  all  $\xi \in ]0,1[$,
$p\ge 1$,  $x,y \in \Z^d$, $\phi,\psi \in \Ran P_{E_{n,\omega},\omega}$, $n\in N$,
 and  $i, j\in \{1,2,\dots,\nu_{n,\omega}\}$  we have
\begin{align}\notag
&\lVert \chi_x\phi  \rVert \lVert \chi_y\psi \rVert  
\le \left[W_{E_{n,\omega},\omega}(x)
 W_{E_{n,\omega},\omega}(y)\right]
\left[\lVert T_x^{-1}  \phi \rVert 
\lVert T_y^{-1}  \psi  \rVert\right] \\
& \quad \le 2^{\kappa }\la x\ra^{\kappa}\la y\ra^{\kappa} 
\lVert T_x^{-1}  \phi \rVert 
\lVert T_y^{-1}  \psi  \rVert
\left[C_{I,\xi,p,\omega} \la y\ra^{2\kappa}
\e^{- |x-y|^\xi} \right]^{\frac 1 p} \label{preSUDEC}\\
& \quad \le C_{I,\xi,p,\omega}^\prime
\lVert T_x^{-1}  \phi \rVert 
\lVert T_y^{-1}  \psi  \rVert
 \la y\ra^{\frac {2(p+1)\kappa}{p}} \e^{-\frac 1 {2p}  |x-y|^\xi}, \notag
\end{align}
where we used \eqref{Tab}.

The SUDEC estimate \eqref{decayphi2222} for given $\eps> 0$ and 
$\zeta \in ]0,1[$ follows from \eqref{preSUDEC} by working with
$\frac d 2 < \kappa < \frac {d+ \eps} 2$,  choosing $p\ge 1$ such that
$d +\eps= \frac {2(p+1)\kappa}{p}$,  and taking $\xi =\frac {1 + \zeta} 2 $.

To prove the SULE-like estimate \eqref{SULE},
for each $n \in \N$ we take a nonzero eigenfunction
 $\psi \in  \Ran P_{E_{n,\omega},\omega}$,
and pick $y_{n,\omega} \in \Z^d$ (not unique) such that
\begin{equation}\label{psi}
\lVert \chi_{y_{n,\omega} }\psi \rVert =
\max_{y \in \Z^d} \,
\lVert \chi_y\psi \rVert .
\end{equation}
Since for all $a \in \Z^d$ and $\phi \in \H$ we have
\begin{equation}
\begin{split}
\lVert T_a^{-1}  {\phi}  \rVert^2& = 
\sum_{y \in \Z^d}  \lVert \chi_yT_a^{-1}{\phi}  \rVert ^2 
\le  \max_{y \in \Z^d} \,
\lVert \chi_y{\phi}  \rVert^2    \sum_{y \in \Z^d}  \lVert \chi_yT_a^{-1}\rVert^2\\
& = \max_{y \in \Z^d} \,
\lVert \chi_y{\phi}  \rVert^2    \sum_{y \in \Z^d}  \lVert \chi_y T^{-1}\rVert^2
\le C_d^2 \max_{y \in \Z^d} \, \lVert \chi_y{\phi}  \rVert^2 ,
\end{split}
\end{equation}
we get
\begin{equation}\label{Tmaxphi}
\lVert T_a^{-1} \psi \rVert \le
 C_d\lVert \chi_{y_{n,\omega} }\psi \rVert \quad \text{for all
 $a \in \Z^d$}.
\end{equation}

It now follows from \eqref{preSUDEC}, taking $\psi$ as in \eqref{psi},
   $y= y_{n,\omega}$,
using  \eqref{Tmaxphi}, and choosing $p$ and $\xi$ as above, that for all 
$x \in \Z^d$, $\psi \in  \Ran P_{E_{n,\omega},\omega}$,
and  $ i\in \{1,2,\dots,\nu_{n,\omega}\}$ we have 
\begin{align}
\lVert \chi_x\phi\rVert \le
 C_d^{-1}  C_{I,\zeta,\eps,\omega}^{\prime\prime}\lVert T^{-1}  \phi  \rVert
 \la  y_{n,\omega}\ra^{d+\eps}
\,  \e^{-|x-y_{n,\omega}|^\zeta},
\end{align}
which is just  \eqref{SULE}.

SUDEC and SULE for  
the complete orthonormal set  
$\{\phi_{n,j,\omega}\}_{ n\in \N, j\in \{1,2,\dots,\nu_{n,\omega}\} }$
of eigenfunctions of $H_\omega$ with energy in $I$ follows.
Note that  the equalities \eqref{sumalphan} and
 \eqref{sumalpha} follow immediately from   \eqref{trestmeas}.

To prove \eqref{NL}, note that it follows from \eqref{SULE5} that
\begin{align}\notag
& \norm{\chi_{\{|x-y_{n,\omega}|\ge R\}}\phi_{n,j,\omega}}^2 \\
& \quad \le   C_{I,\zeta,\eps,\omega}^2  \la y_{n,\omega}\ra^{2(d + \eps)} 
\alpha_{n,j,\omega}  \sum_{x \in \Z^d, |x-y_{n,\omega}|\ge R} \e^{-|x-y_{n,\omega}|^\zeta} \notag\\
&\quad \le  
C_{I,\zeta,\eps,\omega}^\prime  \la y_{n,\omega}\ra^{2(d + \eps)}
 \alpha_{n,j,\omega} \e^{-\frac 1 2 R^\zeta}  \le \tfrac 1 2,
\end{align}
if we take 
\begin{equation}
 R=R_{n,j,\omega} 
\ge 2\left\{ \log  \left(2 C_{I,\zeta,\eps,\omega}^\prime  \la y_{n,\omega}\ra^{2(d + \eps)}
 \alpha_{n,j,\omega}\right)\right\}^{\frac 1 \zeta}.
\end{equation}

Given $L\ge 1$, we set
\begin{equation}
\begin{split}
 R_{L,\omega}& =
\ 2\left\{ \log  \left(2C_{I,\zeta,\eps,\omega}^\prime  \la L\ra^{2(d + \eps)}
 \alpha_{n,j,\omega}\right)\right\}^{\frac 1 \zeta} \le 
C_{I,\zeta,\eps,\omega}^{\prime\prime} \left(\log L\right)^{\frac 1 \zeta} ,\\
 S_{L,\omega}& = L +  2 R_{L,\omega} \le 
C_{I,\zeta,\eps,\omega}^{\prime\prime\prime} L.
\end{split}
\end{equation}
Note that if $ |y_{n,\omega}| \le L$ we have
 $\norm{\chi_{0, S_{L,\omega}}\phi_{n,j,\omega} }^2\ge \frac 12$ for all
$ j\in \{1,2,\dots,\nu_{n,\omega}\}$.  
Thus, using \eqref{cov} and \eqref{trest2}, we get
\begin{align}\notag
& \tfrac 12 N_L  \le \sum_{n \in \N,j\in \{1,2,\dots,\nu_{n,\omega}\} }  
\|\chi_{0, S_{L,\omega}}\phi_{n,j,\omega}\|^2 =
\|\chi_{0, S_{L,\omega}}P_{I,\omega}\|^2_2\\ 
& \quad \le \sum_{a\in \Z^d \cap \Lambda_{S_{L,\omega}}(0)}
\|\chi_{a}P_{I,\omega}\|^2_2 = \sum_{a \in \Z^d \cap \Lambda_{S_{L,\omega}}(0)}
\|\chi_{0}P_{I,\tau(-a) \omega}\|^2_2
\\&\quad \le C_d \sum_{a\in \Z^d \cap \Lambda_{S_{L,\omega}}(0)}
\mu_{\tau(-a) \omega}(I)  \le C^\prime_d S_{L,\omega}^d K_I
\le \tilde{C}_{I,\zeta,\eps,\omega} K_I  L^d,
\notag
\end{align}
which yields \eqref{NL}.
\end{proof}

\section{SUDEC with exponential decay}\label{secexpdec}

In this section we prove Theorem~\ref{thmsudecapp}.

\begin{proof}[Proof of Theorem~\ref{thmsudecapp}]
  Let us fix $\eps>0$. Since $\bar{I} \subset \Xi^{\text{CL}}_\I$, 
we can pick  $\zeta\in ]0,1[$ and $\alpha \in ]1,\zeta^{-1}[$ and
such that   $\alpha < (1+\eps)\zeta$ and  there is a length scale
$L_0\in 6 \mathbb{N}$
 and a mass $m=m_\zeta>0$, so if we
 set 
$L_{k+1} = [L_k^\alpha]_{6\mathbb{N}}$, $k=0,1,\dots$,
 we have \eqref{MSAest} 
for all $k=0,1,\ldots$, and  $x, y \in \mathbb{Z}^d$
with 
$|x-y| > L_k +\varrho$.  We fix  $\rho \in ]\frac 23,1[$ and  $b > \frac{1+2\rho}{
1-2\rho}>1$.   As in  \cite[Proof of Theorem 6.4]{K4}, 
we pick  $\rho \in ]\frac 13,\frac 1 2[$ and  $b > \frac{1+2\rho}{
1-2\rho}>1$, and for each  $x_0 \in  \mathbb{Z}^d$ and
 $k = 0,1,\cdots $
define the discrete annuli
\begin{align}
A_{k+1}(x_0)& =\left\{
 \Lambda_{2bL_{k+1}}(x_0)\setminus \Lambda_{2L_k}(x_0) \right\}
\cap \mathbb{Z}^d  ,\\
\tilde A_{k+1}(x_0) &= \left\{\Lambda_{{\frac{ 2b}{ 1+\rho}}L_{k+1}}(x_0)\setminus
\Lambda_{{\frac 2{1-\rho}}L_k}(x_0)\right\}
\cap \mathbb{Z}^d  .\label{tildeA}
\end{align}

We consider the event
\begin{align} F_k  =  \bigcap_{y\in\Z^d,\, \log  {\la y \ra} \le 
\left(m L_{k+1}\right)^{(1+\eps)^{-1}}}\  \bigcap_{x\in A_{k+1}(y)} R(m,L_k,I,x,y) ,
\end{align} 
with $R(m,L,I,x,y)$ given in \eqref{defsetR}.
It follows from \eqref{MSAest}  that $\sum_{k= 1}^\infty \P(F_k^c) <\infty$, so
that the Borel-Cantelli Lemma applies and yields an almost-surely finite
 $k_1(\omega)$, such that  for  all $k\ge
k_1(\omega)$, if  $E\in I$ and $\log  {\la y \ra} \le 
\left(m L_{k+1}\right)^{(1+\eps)^{-1}}$,  either $\Lambda_{L_k}(y)$ is
$(\omega,m,E)$-regular or  $\Lambda_{L_k}(x)$ is $(\omega,m,E)$-regular for all
$x\in A_k(y)$.  For convenience we require $k_1(\omega)\ge 1$.

Using \cite[Lemma 6.2]{K4} we conclude  that for all $y \in \Z^d$,
 $\P$-a.e. $\omega$, and
$\mu_\omega$-a.e. $\lambda \in \I$,  there exists
a  finite $k_2=k_2(y,\omega,\lambda)$
such that
for all $k > k_2$ we have that  $\Lambda_{L_k}(y)$ is
$(\omega,m,\lambda)$-singular, and moreover  
$\Lambda_{L_{k_2 }}(y)$ is $(\omega,m,\lambda)$-regular unless 
 $k_2(\omega,y,\lambda)=0$.

For each $y \in \Z^d$ we define  $k_3:=k_3(y)$  by
\begin{align}\label{defk3}
\left(m L_{k_3}\right)^{(1+\eps)^{-1}} <\log  {\la y \ra} \le 
\left(m L_{k_3+1}\right)^{(1+\eps)^{-1}},
\end{align}
when possible, with $k_3(y)=-1$ otherwise.

We now set
\begin{equation}
k_\ast:=k_\ast(\omega,y,\lambda)=
\max \{k_1(\omega), k_3(y),k_2(\omega,y,\lambda)+1\};
\end{equation}
note that $1\le k_\ast(\omega,y,\lambda)< \infty$  for
 $\P$-a.e. $\omega$, and
$\mu_\omega$-a.e. $\lambda \in \I$.

Let  $\phi,\psi\in\H_+$ be given.  
Then for  $\P$-a.e. $\omega$, and
$\mu_\omega$-a.e. $\lambda \in \I$,
if $k\ge k_\ast$ the box
$\Lambda_{L_k}(y)$ is $(\omega,m,\lambda)$-singular and thus $\Lambda_{L_k}(x)$ is
$(\omega,m,\lambda)$-regular  for all $x\in A_{k+1}(y)$. It follows,
as in \cite[Proof of Theorem 6.4]{K4}, that
for all $x \in \tilde A_{k+1}(y) $  we have
\begin{align}\label{xreg}
\|\chi_x \Pb_{\omega}(\lambda)\psi\|   \le C_{d,m} 
 \langle y\rangle^{\kappa}\| T^{-1}
\Pb_{\omega}(\lambda)\psi\| \mathrm{e}^{-m_\rho |x-y|} ,
\end{align} 
where  $m_\rho=\frac {\rho(3 \rho -1)}{2} m \in ]0,m[$.
It remains to consider the case when 
 $x \in \Lambda_{{\frac 2{1-\rho}}L_{k_\ast}}(y) \cap \Z^d$.  If
 $k_\ast=\max \{k_1(\omega),k_3(y)\}> k_2(\omega,y,\lambda)$,
 we use
 \eqref{boundGW}  and, 
if  $k_\ast=k_3(y)$, \eqref{defk3}, getting
\begin{align}
&\|\chi_x \Pb_{\omega}(\lambda)\psi\|  \le C_{d} \| T_x^{-1}
\Pb_{\omega}(\lambda)\psi\| 
\mathrm{e}^{m L_{k_\ast}}
\mathrm{e}^{-m L_{k_\ast}} \\ &
\quad  \le
\begin{cases}  C_{d}\la x\ra^\kappa 
\| T^{-1} \Pb_{\omega}(\lambda)\psi\| \mathrm{e}^{({\log \la y\ra})^{1+\eps}}
\mathrm{e}^{-m |x-y|} & \text{if  $k_\ast=k_3(y)$}\\
 C_{d} \la x\ra^\kappa \|T^{-1} \Pb_{\omega}(\lambda)\psi\| 
\mathrm{e}^{m L_{k_1(\omega)}}
\mathrm{e}^{-m |x-y|} & \text{if  $k_\ast=k_1(\omega)$}
\end{cases}   .\notag
\end{align}
Estimating $\|\chi_y \Pb_{\omega}(\lambda)\phi\|$ by
\eqref{boundGW},  we  get the bound
\begin{align}\label{sudecapp}
&\|\chi_x \Pb_{\omega}(\lambda)\psi\|  \|\chi_y \Pb_{\omega}(\lambda)\phi\|\\
& \qquad \qquad 
\le C_{d,\omega}\scal{x}^\kappa \scal{y}^{2\kappa}
\sqrt{\alpha_{\lambda,\phi}\alpha_{\lambda,\psi}}\,
\mathrm{e}^{({\log \la y\ra})^{1+\eps}} \mathrm{e}^{-m^\prime |x-y|},  \notag
\end{align}
with $m^\prime =m_\rho$.
If  $k_\ast=k_2(\omega,y,\lambda)+1> \max \{k_1(\omega),k_3(y)\} $, we must have  $k_2\ge 1$  and 
hence $\Lambda_{L_{k_2 }}(y)$ is $(\omega,m,\lambda)$-regular.
Using \eqref{eigdec9} and   \eqref{Tab}, we get 
\begin{equation}
 \|\chi_y \Pb_{\omega}(\lambda)\phi\|   
 \le C_{d,I,m}
\scal{y}^{\kappa} \| T^{-1} \Pb_{\omega}(\lambda)\phi\| 
\mathrm{e}^{- m \frac{ L_{k_2}} 4} . \label{yk2reg}
 \end{equation}
 If   $x \in \Lambda_{{\frac 2{1- 2\rho }}L_{k_2}}(y) \cap \Z^d$, we may bound the 
term in $x$ by \eqref{boundGW}
and   get \eqref{sudecapp} with $m^\prime =\frac{(1-2\rho ) m}4$
and another
 constant  $C_{d,\omega}$.
Since  $x \in \Lambda_{{\frac 2{1-\rho}}L_{k_2+1}}(y) \cap \Z^d$,
we cannot have $x \notin \Lambda_{\frac {2b}{1+2\rho } L_{k_2+1}}(y) \cap \Z^d$
by our choice of $b$ and $\rho$. 
 Thus the only remaining case is    when
$x \in {\tilde {A}}^\prime_{k_2+1}(y)$, where ${\tilde {A}}^\prime_{k_2+1}(y)$
is defined as in \eqref{tildeA} but with $2 \rho$ substituted for $\rho$.   
If all boxes   $\Lambda_{L_{k_2 }}(x^\prime)$
 with $\lvert x^\prime -x\rvert \le \rho  \lvert x-y \rvert$ 
are $(\omega,m,\lambda)$-regular, the argument in \cite[Proof of Theorem 6.4]{K4}
still applies, and hence we   also get \eqref{xreg} and  
\eqref{sudecapp} with with $m^\prime =m_\rho$.  If not, there exists
$x^{\prime}  \in  \tilde A_{k_2+1}(y)$ with  
$\lvert x^\prime -x\rvert \le  \rho  \lvert x-y \rvert$ such that
 $\Lambda_{L_{k_2 }}(x^\prime)$ is $(\omega,m,\lambda)$-singular.  Clearly,
 $x^\prime \in  \tilde A_{k_2+1}(y)$
if and only if $y \in  \tilde A_{k_2+1}(x^\prime)$.  In addition, since 
$k_3(y)\le  k_2(\omega,y,\lambda) $
we have  $k_3(x^\prime) \le  k_2(\omega,y,\lambda)+1$, as
\begin{equation}
\log {\la x^\prime\ra}\le \tfrac 1 2 \log 2 + \log {\la y\ra} + 
 \log {\la b L_{k_2+1}\ra} \le
 \left(m L_{k_2+1}\right)^{(1+\eps)^{-1}}.
\end{equation}
Thus,  as $k_2\ge k_1(\omega)$, we can  apply the argument leading to 
\eqref{xreg} in the annulus $\tilde A_{k_2+1}(x^\prime)$, obtaining
\begin{align}\label{xreg4}
\|\chi_y \Pb_{\omega}(\lambda)\phi\| &  \le C_{d,m} 
 \langle x^\prime\rangle^{\kappa}\| T^{-1}
\Pb_{\omega}(\lambda)\phi\| \mathrm{e}^{-m_\rho |x^\prime-y|}\\
& \le  C^\prime_{d,m} 
 \langle y \rangle^{\kappa}\| T^{-1}
\Pb_{\omega}(\lambda)\phi\| \mathrm{e}^{-\rho(1-\rho) m_\rho |x-y|}
 ,
\end{align} 
where we used $ |x^\prime-x| \le \rho  |x-y|$
and   $ |x^\prime-y| \ge |x-y| -  |x^\prime-x| \ge (1-\rho) |x-y|$.
Estimating $\|\chi_x \Pb_{\omega}(\lambda)\psi\|$ by
\eqref{boundGW},  we  get the bound
\begin{align}\label{sudecapp5}
\|\chi_x \Pb_{\omega}(\lambda)\psi\|  \|\chi_y \Pb_{\omega}(\lambda)\phi\|
\le C_{d,\omega}\scal{x}^\kappa \scal{y}^{\kappa}
\sqrt{\alpha_{\lambda,\phi}\alpha_{\lambda,\psi}}\,
\mathrm{e}^{-m^\prime |x-y|} 
\end{align}
with $m^\prime =\rho(1-\rho)m_\rho$.

The thorem is proved.
\end{proof}

\section{Decay of the Fermi projection}\label{secproofFermi}

In this section we prove Theorem~\ref{thmfermi}.

\begin{proof}[Proof of Theorem~\ref{thmfermi}]
Let $I $ and $I_1$ be
 bounded open intervals with $\bar{I} \subset I_1\subset \bar{I_1} \subset  
\Xi^{\text{CL}}_\I$. 
It follows from \cite[Theorem 3.8]{GK1} 
that for  all  $\zeta\in ]0,1[$ we have
\begin{equation} \label{thm38}
\E \left\{\sup_{f \in\mathcal{B}_1 } 
\left\|\chi_{x}f(H_\omega) P_\omega(I_1)
\chi_{y}\right\|_2^2\right\}
 \leq   C_{I_1,\zeta }\,\mathrm{e}^{-|x-y|^\zeta} \quad 
\text{for all $x,y \in \mathbb{Z}^d$}.
\end{equation}

We write $I = (\alpha,\beta)$, and  fix $\delta =\frac 12  \mathrm{dist}( I, \partial I_1)>0$. Given $\zeta \in ]0,1[$,
we choose $\zeta^\prime \in ]\zeta,1[$.  Since $H_\omega$ is semibounded, we can choose $\gamma > -\infty$ such that
$\Sigma \subset ]\gamma,\infty[$.  We pick a
 ${\rm L}^1$-Gevrey function $g$  of class $\frac 1 {\zeta^\prime}$ on 
$ ]\gamma,\infty[$, such that  $0\le g \le 1$,  $g\equiv1$ on $]-\infty,\alpha-\delta]$
and $g\equiv0$ on $]\beta+\delta,\infty[$. (See
\cite[Definition 1.1]{BGK}; such a function always exists.)  For all $E \in I$ we have
$P_{\omega}^{(E)}=  g(H_\omega) + f_E(H_\omega)$, where
$f_E(t) = \chi_{]-\infty,E]}(t) - g(t)\in  \mathcal{B}_1$, with   
$f_E(H_\omega) =f_E(H_\omega) P_\omega(I_1)$. Using \cite[Theorem 1.4]{BGK},
 for $\P$-a.e. $\omega$ we have 
\begin{equation}\label{BGKest1}
\left\|\chi_{x}g (H_\omega) \chi_{y}\right\| \le C_{g,\zeta,\zeta^\prime}
\,\mathrm{e}^{-C_{g,\zeta,\zeta^\prime}|x-y|^\zeta}\quad 
\text{for all $x,y \in \mathbb{Z}^d$}.
\end{equation}
On the other hand, it follows from  \cite[Eq. (2.36)]{GK1} and the covariance
\eqref{cov} that for $\P$-a.e. $\omega$
\begin{equation}
\left\|\chi_{x}g (H_\omega) \chi_{y}\right\|_1 \le
 \left\|\chi_{x}g (H_\omega)\chi_x\right\|_1^{\frac 1 2}
 \left\|\chi_{y}g (H_\omega)\chi_y\right\|_1^{\frac 1 2}\le C_{g}
\quad \text{for all $x,y \in \mathbb{Z}^d$}.
\end{equation}
Since $\norm{A}_2^2 \le \norm{A} \norm{A}_1$ for any operator $A$, we get
\begin{equation}\label{BGKest}
\left\|\chi_{x}g (H_\omega) \chi_{y}\right\|_2^2 \le C^\prime_{g,\zeta,\zeta^\prime}
\,\mathrm{e}^{-C^\prime_{g,\zeta,\zeta^\prime}|x-y|^\zeta}\quad 
\text{for all $x,y \in \mathbb{Z}^d$}.
\end{equation}
 
The estimate \eqref{fermidecay} for all $\zeta \in ]0,1[$ now follows from 
\eqref{thm38} and \eqref{BGKest}.

To prove the converse, let us suppose \eqref{fermidecay} holds for some
 $\zeta \in ]0,1[$.)
Let $\mathcal{X}\in C^\infty_{c,+} ({I})$.  By  the spectral theorem,
\begin{align}\notag
\mathrm{e}^{-i tH_{\omega} } \mathcal{X}(H_{\omega})
&= \int {\mathrm{e}^{-i tE} }\mathcal{X}(E)  P_\omega(\di E)
= -\int \left({\mathrm{e}^{-i tE} }\mathcal{X}(E)\right)^\prime   P_{\omega}^{(E)} \di E\\
& =   -\int_I \left({\mathrm{e}^{-i tE} }\mathcal{X}(E)\right)^\prime 
P_{\omega}^{(E)}  \di E .
\end{align}
Thus for all $n>0$ we have
\begin{align}
 \left\| {\langle} x{\rangle}^{\frac n2}
{\mathrm{e}^{-i tH_{\omega} }}\mathcal{X}(H_{\omega})\chi_0 
 \right\|_2 \le C_\mathcal{X} (1 + t) \int_I  \left\| {\langle} x {\rangle}^{\frac n2}
 P_{\omega}^{(E)}\chi_0  \right\|_2 \di E,
\end{align}
and hence 
\begin{align}\notag
&\E \left\{\left\| {\langle} x {\rangle}^{\frac n2}
{\mathrm{e}^{-i tH_{\omega} }}\mathcal{X}(H_{\omega})\chi_0 
 \right\|_2^2\right\} \\
& \quad \le C_\mathcal{X}^2 (1 + t)^2 \E \left\{\left\{\int_I
 \left\| {\langle} x {\rangle}^{\frac n2}
  P_{\omega}^{(E)}\chi_0 
 \right\|_2 \di E\right\}^2\right\} \\
& \quad \le  C_\mathcal{X}^2 (1 + t)^2 |I| \int_I
\E \left\{\left\| {\langle} x {\rangle}^{\frac n2}
 P_{\omega}^{(E)}\chi_0 
 \right\|_2^2 \right\}\di E \le C_{ \mathcal{X},I,n,\zeta} (1 + t)^2,\notag
\end{align}
where we used \eqref{fermidecay} to get the last inequality.  
It follows that 
\begin{align} \notag  
\mathcal{M}(n,\mathcal{X},T)& : =   
\frac2{T} \int_0^{\infty}{\mathrm{e}^{-\frac{2t}{T}}} 
\E \left\{\left\| {\langle} x {\rangle}^{\frac n2}
{\mathrm{e}^{-i tH_{\omega} }}\mathcal{X}(H_{\omega})\chi_0 
 \right\|_2^2\right\}  \di t \\
& \le C_{ \mathcal{X},I,n,\zeta}^\prime (1 + T^2),\label{tam}
\end{align}
 hence
\begin{equation}\label{eqhyp}
\liminf_{T \to \infty}\frac1{T^{\alpha}}
 {\mathcal{M}}(n,\mathcal{X},T)< \infty \quad 
\text{for all $\alpha \ge 2$ and $n >0$}.
\end{equation}
It now follows from  \cite[Theorem~2.11]{GK5} that $I \subset \Xi^{\text{CL}} _\I$.
\end{proof}


\begin{acknowledgement}
F.G. is currently visiting the Universit\'e de Paris Nord with support from the CNRS.

 A.K. was supported in part   by NSF Grant
DMS-0200710. 
\end{acknowledgement}


\end{document}